\documentclass[preprint,12pt]{elsarticle}
\usepackage{graphicx}
\usepackage{graphics}
\usepackage{amsmath}
\usepackage{dcolumn}
\usepackage{amssymb}
\usepackage{bm}

\journal{Annals of Physics}

\begin{document}

\begin{frontmatter}



\title{Kaluza-Klein description of geometric phases in graphene}

\author{K. Bakke}
\ead{kbakke@fisica.ufpb.br} 
\address{Departamento de F\'{\i}sica, Universidade Federal da Para\'{\i}ba, Caixa Postal 5008, 58051-970, Jo\~ao Pessoa, PB, Brazil}
 
\author{A. Yu. Petrov}
\ead{petrov@fisica.ufpb.br}
\address{Departamento de F\'{\i}sica, Universidade Federal da Para\'{\i}ba, Caixa Postal 5008, 58051-970, Jo\~ao Pessoa, PB, Brazil}

\author{C. Furtado}
\ead{furtado@fisica.ufpb.br}
\address{Departamento de F\'{\i}sica, Universidade Federal da Para\'{\i}ba, Caixa Postal 5008, 58051-970, Jo\~ao Pessoa, PB, Brazil}

\begin{abstract}
In this paper, we use the Kaluza-Klein approach to describe topological defects in a graphene layer. Using this approach, we propose a geometric model allowing to discuss the quantum flux in $K$-spin subspace. Within this model, the graphene layer with a topological defect is described by a four-dimensional metric, where the deformation produced by the topological defect is introduced via the three-dimensional part of metric tensor, while an Abelian gauge field is introduced via an extra dimension. We use this new geometric model to discuss the arising of topological quantum phases in a graphene layer with a topological defect.
\end{abstract}

\begin{keyword}
Geometric Phase\sep Graphene \sep Kaluza-Klein Theory
\end{keyword}

\end{frontmatter}

% \linenumbers


\section{Introduction}

The concept of the extra dimensions, which is treated now as one of the most popular concepts in the high energy physics, has a long story. Initially \cite{kk1,kk2}, the hypothesis of extra dimensions has been introduced as an attempt to unify gravity and electromagnetism. Despite this attempt had not succeeded and was abandoned, the idea of introducing additional spacetime dimensions as a possible instrument allowing to construct unified physical theories had found widest applications in the quantum field theory, especially due to the development of the string theory, which is well-known to be consistent namely in the space with extra dimensions \cite{GSH}.

The essence of the Kaluza-Klein approach in its modern form looks like follows: let us suggest that the spacetime, besides of the usual dimensions, involves compact extra dimensions. As a consequence, all fields defined in this space can be expanded in Fourier series with respect to the extra coordinates. Different Fourier modes arising throughout this expansion possess different masses depending on the size of the compact extra dimensions, thus, in principle, the infinite tower of new particles with infinite spectrum of masses can arise. Beside of the string applications, the Kaluza-Klein approach turned out to be an efficient description of black holes (for a review on Kaluza-Klein theory see \cite{SS}). 

The graphene, whose different aspects have been studied earlier in a number of papers \cite{Ja,sit2,Voz,voz3,voz4}, is a physical system whose description essentially involves topological defects. Different aspects of the graphene have been studied with use of the gauge fields approach \cite{Voz,voz3,voz4}, finite temperature methods \cite{Sit}, Berry phases from the presence of dislocations in the graphene layer \cite{mes}, and the parallel transport of Dirac fermions in the presence of a torsion and a curvature \cite{mes2}. Recently, several investigations have established {the physical similarity between gravity and some models used in condensed matter theory} \cite{kat,kleinert,moraesG2}. It is well-known that smooth deformations of graphene sheets produce a gauge field similar to the electromagnetic one \cite{moro,mopu}, and topological defects in graphene can be interpreted as a source of a non-Abelian gauge field \cite{voz1,voz2,lc1,lc2}. Basing on this bridge between the physics of graphene and properties of the gauge and gravitational fields, in this work, we use the Kaluza-Klein theory to describe a graphene layer with a topological defect. The success of applying the quantum field theory concepts in the condensed matter theory allows to hope that the using the Kaluza-Klein theory for describing certain condensed matter models could be very efficient. In particular, the very promising idea consists in using the Kaluza-Klein approach to graphene, where the extra dimension concept naturally emerges in the case of the presence of the topological defect. 

In this paper, we extend the two-dimensional metric which describes a topological defect in a graphene layer to a three-dimensional metric by adding an extra compact dimension to describe the Fermi-point degree of freedom, and demonstrate that we can calculate the geometric phase acquired by the wave function of a massless fermion which arises from the mix of the Fermi points, and the parallel transport of a spinor around the apex of a defect in a graphene layer.

Geometric quantum phases is a term introduced by Berry \cite{berry} to describe the phase shifts acquired by the wave function of a quantum particle in a adiabatic evolution. A well-known quantum phase that belongs to this more general class of phase is the Aharonov-Bohm effect \cite{ab}. Within this effect, the wave function of the electron acquires a topological quantum phase circulating a solenoid. The electron moves in a region where the magnetic field is zero, and still feels the influence of the magnetic field via the vector potential in the phase acquired by the wave function describing the motion of the electron. Aharonov and Anandan \cite{ab1} extended the study of geometric quantum phases for any cyclic evolution, and the phase shift associated with any cyclic evolution is known as the Aharonov-Anandan quantum phase. The study of geometric phases in quantum system has attracted a great deal of attention in recent years \cite{tg1,comp}, with the most important quantum effect is the Aharonov-Bohm effect \cite{ab}. In graphene, the geometric phase arising from the presence of a disclination in a graphene layer has been pointed out in \cite{lc1,lc2}. In this paper, we extend the two-dimensional metric which describes a topological defect in a graphene layer to a three-dimensional metric by adding an extra compact dimension to describe the Fermi-point degree of freedom, and demonstrate that we can calculate the geometric phase acquired by the wave function of a massless fermion which arises from the mix of the Fermi points and the parallel transport of a spinor around the apex of a defect in a graphene layer. 

This paper is organized as follows. In section II, we a make a brief review of graphene, and the appearance of quantum fluxes due to the presence of a topological defect. In section III, we introduce the concept of extra dimension from the Kaluza-Klein theory \cite{kk1,kk2} in graphene, and discuss how to obtain the quantum flux from the $K$-spin part of graphene in a geometrical point of view. Finally, in section IV, we present our conclusions.

\section{Graphene: a brief review}

\begin{figure}
\includegraphics[width=10cm]{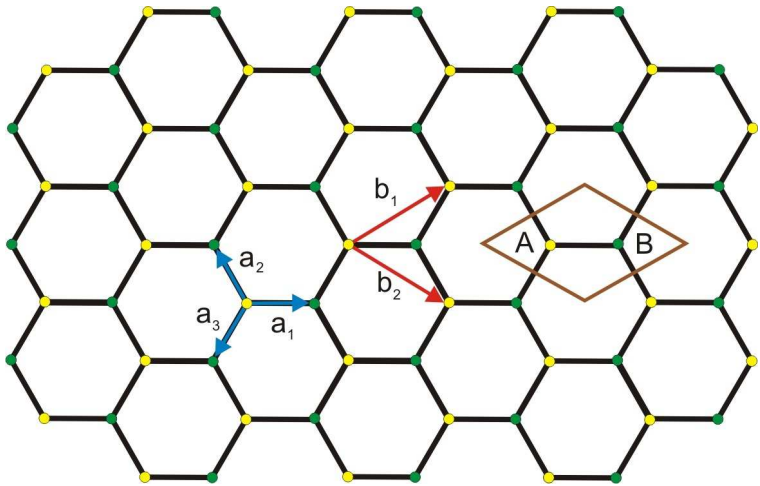}
\caption{A graphene honeycomb lattice described by a yellow and green sites corresponding to the $A$ and $B$ sublattices. The vectors $\bold{b_{1}}$ and  $\bold{b_{2}}$ are vector of the unitary cell. A choice of the unitary cell are represented in right by polygon containing the sites $A$ and $B$. The vectors $\bold{a_{i}}$ represents the nearest neighbors of the lattice}
\label{fig1}
\end{figure}

In this section, we make a brief review of graphene and the arising of quantum phases caused by the presence of a topological defect called disclination \cite{kat,kleinert,moraesG2,furt,staro}.  The band of conduction of the graphene, which consists of a graphite layer, can be described by the tight-binding model \cite{voz1,voz2,lc1,lc2}. A graphene layer is a two-dimensional material formed by a isolated layer of carbon atoms arranged in a honeycomb lattice. We describe this structure by two sublattices $A$ and $B$, where an unitary cell and the vector of the unitary cell are represented in the Figure \ref{fig1}. 

By considering only nearest neighbour interactions in the tight-binding approximation, which includes $\pi$-orbitals perpendicular to the plane at each $C$ atom, the hopping of an electron on a planar honeycomb lattice is described by the following  Hamiltonian \cite{prb:div,prl:seme,ps1}
\begin{equation}\label{tb}
H=-t\sum_{<i,j>}(a^{\dagger}_{i}a_{j} +a_{i}a^{\dagger}_{j}),
\end{equation}
where the operator $a^{\dagger}_{i}$ creates an electron at a site $i$, $a_{j} $ annihilates it at a site $j$, and $t$ corresponds to the hopping amplitude. Considering the periodic properties of the graphene structure, we can diagonalize the Hamiltonian (\ref{tb}) by using the Fourier transform, which reduces the problem of finding the values of energy to the problem of the diagonalization of the Hamiltonian in a single unit cell. Indeed, for $a(\vec{k})=e^{\imath \vec{k}\cdot\vec{i}}a_{i}$, we obtain the following expression for the Hamiltonian:
\begin{eqnarray}\label{matrixH}
H=-t\int d^{2}k(a^{\dagger}_{\mathcal{A}}(\vec{k})a^{\dagger}_{\mathcal{B}}(\vec{k}))\left(
\begin{array}{cc}
0 & \sum^{3}_{i=1}e^{\imath \vec{k}\cdot \vec{v}_{i}}\\
\sum^{3}_{i=1}e^{-\imath \vec{k}\cdot \vec{v}_{i}}& 0 \\ 
\end{array} \right)\left( \begin{array}{c}
a_{\mathcal{A}}(\vec{k})\\
a_{\mathcal{B}}(\vec{k})
\end{array}\right), 
\end{eqnarray}
where the Fourier transformed of operators $a_{\mathcal{A}}(\vec{k})$ and $a_{\mathcal{B}}(\vec{k})$ are associated with of the sublattice $\mathcal{A}$ and $\mathcal{B}$. We can diagonalize the Hamiltonian (\ref{matrixH}) and obtain the following expression for the values of energy
\begin{eqnarray}
\label{eingen}
E=\pm t\sqrt{ 1+ 4 \cos ^{2}\frac{\sqrt{3}k_{y}}{2} +4 \cos \frac{3k_{x}}{2}\cos \frac{\sqrt{3}k_{y}}{2}},
\end{eqnarray}
where the interatomic distance is normalized to one. For special isolated values of $k$, the eigenvalues of the energy (\ref{eingen}) are zero. These zero points are known as Fermi points or $K$-points. From the relation (\ref{eingen}) we obtain six $K$-points, where two of them  are inequivalent. We select the vectors $\vec{K}_{+}=\frac{2\pi}{3}(1, \frac{1}{\sqrt{3}})$ and $\vec{K}_{-}=-\frac{2\pi}{3}(1, \frac{1}{\sqrt{3}})$. The structure band of graphene consists of an approximately empty conduction band and a completely filled valence band. In this case, the behavior of graphene is dominated by the behavior of the Hamiltonian (\ref{matrixH}) near the Fermi points. Therefore we can develop a continuous theory for graphene near the Fermi points. A continuous theory for graphene based on the tight-binding theory describing both local and global properties is presented in Refs. \cite{prb:div,prl:seme}. These continuous models complement atomistic methods incorporating long range effects, which are dificult to derive from the first principles. In this context, the Fermi surface reduces to two points ($K$-points) located at the Brillouin zone. In the low-energy limit, the properties of graphene can be described by the free fermions theory \cite{prb:div} within a continuum model based on the Dirac equation \cite{voz1,voz2}. Namely, in the absence of interactions (and in the low-energy limit), the Hamiltonian of the system looks like
\begin{eqnarray}
H_{0}=-i\hbar\,v_{f}\,\left(\sigma^{x}\partial_{x}+\sigma^{y}\partial_{y}\right).
\label{2.1}
\end{eqnarray}
Here, $\sigma^{i}$ correspods to the Pauli spin matrices (acting on the $\mathcal{A}/\mathcal{B}$ labels), and $v_{f}$ corresponds to the Fermi velocity. The corresponding states of this system are labeled by the direction of the wave vector $\textbf{K}$ and by the index $\mathcal{A}$ or $\mathcal{B}$ for the sublattices. We refer to these states as to $\left|\mathbf{K}_{\pm},\mathcal{A}\right\rangle$ and $\left|\mathbf{K}_{\pm},\mathcal{B}\right\rangle$. Since the Fermi level space is a four-dimensional space, we can choose these states as a basis by considering $\mathbf{K}_{-}=-\mathbf{K}_{+}$. Thus, the Hamiltonian (\ref{2.1}) describes the transfer of an electron only from the $\mathcal{A}$ sublattice to the $\mathcal{B}$ sublattice and \textit{vice versa}.

Recently, the influence of topological defects in graphene layers on the transport properties of graphene has been investigated in a number of papers \cite{voz1,voz2,lc1,lc2}. Topological defects in graphene (disclinations, edge dislocations, etc.) can be conceptually generated by a ``cut and glue'' process which is known in the literature as the Volterra process \cite{kat,kleinert,moraesG2}. Recently, several authors have investigated geometric phases in graphene in the presence of topological defects. In Ref. \cite{fur}, one of us has used the parallel transport to investigated geometric phase in graphene by using the geometric theory of defects. When we transport a spinor $\Psi$ around of defect in a graphene layer with a defect, we obtain
\begin{eqnarray}
\Psi'=U(C)\Psi,
\label{holonomy}
\end{eqnarray}
where $U(C)$ provides the geometric phase obtained by the spinor in this process. The matrix $U(C)$ is the holonomy matrix, and a non-trivial holonomy can be interpreted as the analogue of the Aharonov-Bohm effect, where the magnetic flux is replaced by a ``geometric flux''. For example, disclinations in a graphene layer can be generated via the Volterra process when we remove or insert a $\pi/3$ sector in a honeycomb lattice. As a result, afterwards, the hexagon in the lattice is replaced by a pentagon or heptagon. By cutting a $\pi/3$ sector and gluing the opposite sides (as it is done, e.g., in \cite{lc2,ps1}), it can be described by using two equations which can be interpreted in terms of the fluxes of fictitious gauge fields through the apex of the cone: 
\begin{eqnarray}
\oint\omega_{\mu}\,dx^{\mu}=-\frac{\pi}{6}\,\sigma^{z};\,\,\,\,\,\,\oint\ A_{\mu}\,dx^{\mu}=\frac{\pi}{2}\,\tau^{y}.
\label{2.2a}
\end{eqnarray}

The first equation of (\ref{2.2a}) measures the angular deficit of the cones arising when a spinor is transported in a parallel way around the apex of the cone produced by the variation of the local reference frame where a spinor is defined \cite{weinberg}. The general expression of the parallel transport of a spinor around the apex of the cone in a graphene layer is given by a unitary transformation called holonomy \cite{fur,bfs}. 
\begin{eqnarray}
U_{1}\left(C\right)=\mathcal{P}\exp\left[-\oint\Gamma_{\mu}\left(x\right)\,dx^{\mu}\right]=\exp\left[\frac{1}{2}\oint\left(1-\frac{n_{\Omega}}{6}\right)\sigma^{3}\,d\varphi\right],
\label{2.2b}
\end{eqnarray}
where $\Gamma_{\mu}\left(x\right)$ corresponds to the spinorial connection \cite{weinberg}, and $n_{\Omega}$ corresponds to the number of sectors which can be removed or inserted into a graphene layer.

The second equation of (\ref{2.2a}) corresponds to the flux of the $K$-spin flux, which corresponds to the mix of the Fermi points $\textbf{K}_{+}$ and $\textbf{K}_{-}$ when one $\pi/3$ sector is removed from the graphene layer \cite{lc1,lc2}.  This cut (introduced via the Volterra process) yields an exchange between the $\mathcal{A}$ and $\mathcal{B}$ sublattices causing an apparent discontinuity when the spinor is transported around the defect.  But, the parallel transport of the spinor along the cut must be smooth. To compensate this fact, we should introduce gauge fields that remove this fictitious discontinuity. Futhermore, the $\tau^{i}$ matrices which are present in the second equation of (\ref{2.2a}) correspond to the standard Pauli matrices acting only on the Fermi-points indices $\pm$. The general expression for the holonomy associated with non-Abelian gauge field $A_{\mu}$ in the second equation of (\ref{2.2a}) can be written in the form \cite{lc1}:
\begin{eqnarray}
U_{2}\left(C\right)=\mathcal{P}\exp\left[-\oint A_{\mu}\left(x\right)\,dx^{\mu}\right]=\exp\left[-i\oint\frac{n_{\Omega}}{4}\,\tau^{y}\,d\varphi\right].
\label{2.2c}
\end{eqnarray}
It is worth to mention that if we remove or insert an even number of $\pi/3$ sectors, the second quantum phase in (\ref{2.2a}) becomes equal to $\pm1$. This happens because the Fermi-points $\mathbf{K}_{\pm}$ are not mixed \cite{lc1,lc2}. Up to now, the non-abelian gauge field $A_{\mu}$ in the second equation of (\ref{2.2a}) is introduced to describe the quantum flux from the mix of the Fermi points and has no geometrical description. In the next section, we shall see, with the description of the $K$-spin based on an extra dimension of the Kaluza-Klein theory \cite{kk1,kk2,kk3}, that the $K$-spin enters into the Dirac equation as a geometrical quantity. The four-dimensional Fermi level space $\left(\left|\mathbf{K}_{+},\mathcal{A}\right\rangle,\,\left|\mathbf{K}_{+},\mathcal{B}\right\rangle,\,\left|\mathbf{K}_{-},\mathcal{A}\right\rangle,\,\left|\mathbf{K}_{-},\mathcal{B}\right\rangle\right)$ will be represented by usual Dirac 4-component spinors.

%%%%%%%%%%%%%%%%%%%%%%%%%%%%%%%%%%

\section{Extra dimensions approach in a graphene layer}

We begin this section with a brief review of the Kaluza-Klein theory \cite{kk1,kk2,kk3}. In the following, we show how the Kaluza-Klein theory can be used to describe the Fermi-point degree of freedom ($K$-spin) from the a geometrical point of view. It has been shown in Ref. \cite{fur} that the first quantum flux in (\ref{2.2a}) can be obtained within the geometrical approach: a holonomy matrix is constructed when the spinor suffers a parallel transport around the apex of a graphitic cone. The effect of the parallel transport of a spinor around the apex of the graphitic cone is analogous to the Aharonov-Bohm effect \cite{ab}. However, a geometrical description for the second quantum flux in (\ref{2.2a}) is not known yet. In this work, we suggest to give this description via the Kaluza-Klein theory \cite{kk1,kk2,kk3}, \textit{i.e.}, we intend to describe the Fermi-point degree of freedom or $K$-spin via the Kaluza-Klein approach. In the Kaluza-Klein theory, the line element in the presence of a non-Abelian gauge field is given in the form \cite{kk3}:
\begin{eqnarray}
ds^{2}=g_{\mu\nu}\left(x\right)\,dx^{\mu}\,dx^{\nu}+\left(dy+\kappa\,B_{\mu}\,dx^{\mu}\right)^{2},
\label{2.3a}
\end{eqnarray}
where $g_{\mu\nu}\left(x\right)$ is the metric tensor in $\left(2+1\right)$ dimensions, and $y$ corresponds to the coordinate of the extra dimension (the $\kappa$ is a constant called Kaluza constant). Following the prescription of Ref. \cite{kk3}, an Abelian gauge group arises from a coordinate transformation $y\rightarrow y+\xi\epsilon\left(x\right)$, where $\xi$ is a scale factor and $\epsilon\left(x\right)$ is the parameter of the gauge transformation. Therefore, an Abelian gauge field is chosen as $B_{\mu}=\xi\,A_{\mu}\left(x\right)$, where $A_{\mu}\left(x\right)$ corresponds to a gauge field which does not depend on $y$. Based on this coordinate transformation, we have that the off-diagonal elements of metric (\ref{2.3a}) are given by $A_{\mu}\left(x\right)\rightarrow A_{\mu}'\left(x\right)=A_{\mu}\left(x\right)+\partial_{\mu}\epsilon\left(x\right)$. In the continuum approximation of the conical graphene leaf, we introduce the extra dimension $y$ in the two-dimensional metric as
\begin{eqnarray}
ds^{2}=-dt^{2}+d\rho^{2}+\eta^{2}\rho^{2}\,d\varphi^{2}+\left(dy+\frac{n_{\Omega}}{2}\,\rho\,d\varphi\right)^{2},
\label{1.1}
\end{eqnarray}
where $\eta$ is a parameter related to the deficit or excess angle corresponding to the angular sector which is removed or inserted to form the defect. Without the extra dimension, the element (\ref{1.1}) describes the cosmic string spacetime in (2+1) dimensions. Recently, the influence of the topology of the cosmic string spacetime background has been considered within several contexts. For instance, the gravitational analogue of the Aharonov-Bohm effect \cite{val}, geometric quantum phases \cite{bf4}, application of the self-adjoint extension \cite{cm}, classical holonomies with the Fermi-Walker transport \cite{bf15}, and relativistic Einstein-Podolsky-Rosen correlations \cite{bf17}. In the condensed matter physics, the presence of this deficit of angle in a solid corresponds to a linear topological defect called disclination \cite{kleinert,kat,moraesG2,furt}. An interesting discussion about the influence of a disclination on the electronic properties of graphene is presented in Ref. \cite{sit2}. We can also relate the deficit or excess angle to the number of sectors removed from or inserted into the graphite monolayer in the following way:
\begin{eqnarray}
\eta=1-\frac{n_{\Omega}}{6}, 
\label{eta}
\end{eqnarray}
In this expression, the values of $\eta$ in the interval $0\,<\,\eta\,<\,1$ correspond to removing a sector from the leaf to form the defect, $1<\,
\eta\,<\infty$ correspond to inserting a sector into a leaf to form the defect, and the integer number $n_{\Omega}$ is the number of sectors in the graphene layer which can be removed or inserted to the construction of the cones of graphene. The curvature tensor corresponding to the metric (\ref{1.1}) is $R_{\rho,\varphi}^{\rho,\varphi}=\frac{1-\eta}{4\eta}\,\delta_{2}(\vec{r})$, where $\delta_{2}(\vec{r})$ is the two-dimensional delta function. The behavior of the curvature tensor is denominated as a conical singularity \cite{staro,furt}, because it gives rise to the curvature concentrated on the cosmic string axis, while the curvature in all other points of the spacetime is zero. This extra dimension $y$ serves for the description of the $K$-spin space in the continuum approximation of graphene. In (\ref{1.1}), we have that the field $B_{a}$ corresponds to the gauge field that couples the conical geometry of the graphene with the $K$-spin gauge group. In that way, the expression for the only non-zero component of the gauge field $B_{a}$ must be
\begin{eqnarray}
B_{2}=\frac{n_{\Omega}}{2}.
\label{1.2}
\end{eqnarray}
while $\kappa=1$.

Hence, if we want to study the dynamics of massless fermions in a graphene layer, we must define the spinors in the form similar to that one used in quantum field theory in curved spacetime \cite{bd}. Spinors in curved spacetime background are defined in the local reference frame through the components of the noncoordinate basis $\hat{\theta}^{a}=e^{a}_{\,\,\,\mu}\left(x\right)\,dx^{\mu}$, where the components $e^{a}_{\,\,\,\mu}\left(x\right)$ are called tetrads \cite{weinberg}. In the Kaluza-Klein theory, the noncoordinate basis is also written as $\hat{\theta}^{a}=e^{a}_{\,\,\,\mu}\left(x\right)\,dx^{\mu}$, where the Greek indices indicate the spacetime indices ($\mu=t,\rho,\varphi,y$, where $y$ is the extra spatial dimension) and the Latin indices $a=0,1,2,3$ indicates the local reference frame (in particular, the index $a=3$ denotes the extra spatial dimension). The components of this noncoordinate basis satisfy the relation
\begin{eqnarray}
g_{\mu\nu}\left(x\right)=e^{a}_{\,\,\,\mu}\left(x\right)\,e^{b}_{\,\,\,\nu}\left(x\right)\,\eta_{ab},
\label{1.4}
\end{eqnarray}
where $\eta_{ab}=\mathrm{diag}\left(-1,1,1,1\right)$ is the Minkowski tensor. Let us establish that $\hat{\theta}^{0}=dt$, $\hat{\theta}^{1}=d\rho$, $\hat{\theta}^{2}=\eta\rho\,d\varphi$ and $\hat{\theta}^{3}=\frac{n_{\Omega}}{2}\,\rho\,d\varphi+dy$, then, we can write the tetrads in the form: 
\begin{eqnarray}
e^{a}_{\,\,\,\mu}\left(x\right)=\left(
\begin{array}{cccc}
1 & 0 & 0 & 0\\
0 & 1 & 0 & 0\\
0 & 0 & \eta\rho & 0 \\
0 & 0 & \frac{n_{\Omega}}{2}\,\rho & 1\\	
\end{array}\right),\,\,\,\,\,\,
e^{\mu}_{\,\,\,a}\left(x\right)=\left(
\begin{array}{cccc}
1 & 0 & 0 & 0\\
0 & 1 & 0 & 0\\
0 & 0 & \frac{1}{\eta\rho} & 0 \\
0 & 0 & -\frac{n_{\Omega}}{2\eta} & 1\\	
\end{array}\right).
\label{1.7}
\end{eqnarray}

Now, let us begin to to discuss the Dirac equation for a massless fermion in the Kaluza-Klein theory. The general expression for the Dirac equation in this background becomes
\begin{eqnarray}
i\,\gamma^{\mu}\,\left[\partial_{\mu}+\frac{i}{4}\omega_{\mu ab}\left(x\right)\,\Sigma^{ab}\right]\Psi=0,
\label{8}
\end{eqnarray}
where $\Sigma^{ab}=\frac{i}{2}\left[\gamma^{a},\gamma^{b}\right]$. The Dirac matrices $\gamma^{a}$ satisfy the relation $\left\{\gamma^{a},\gamma^{b}\right\}=-2\eta^{ab}$, that is, the $\gamma^{a}$ matrices are defined in the local reference frame and correspond to the standard Dirac matrices given in the Minkowski spacetime. Note that the introduction of an extra dimension through the Kaluza-Klein theory allows us to describe graphene in $\left(3+1\right)$ dimensions. In $\left(3+1\right)$ dimensions, the Dirac spinors are $4$-component spinors and the $\gamma^{a}$ matrices are $4\times4$ matrices. It has already been discussed in Refs. \cite{graf,Ja} that we can build the $4\times4$ Dirac matrices from the $2\times2$ matrices $\tau^{i}$ and $\sigma^{i}$ that acts on the Fermi-points indices and $\mathcal{A}/\mathcal{B}$ labels, respectively:
\begin{eqnarray}
\gamma^{0}&=&\tau^{1}\otimes I=\left(
\begin{array}{cc}
0 & I \\
I & 0 \\
\end{array}\right);\,\,\,\,\,\,
\gamma^{i}=-i\tau^{2}\otimes\sigma^{i}=\left(
\begin{array}{cc}
0 & -\sigma^{i} \\
\sigma^{i} & 0 \\
\end{array}\right);\nonumber\\
[-2mm]\label{8a}\\[-2mm]
\gamma^{5}&=&i\gamma^{0}\gamma^{1}\gamma^{2}\gamma^{3}=\tau^{3}\otimes I=\left(
\begin{array}{cc}
I & 0 \\
0 & -I \\	
\end{array}\right);\,\,\,\,\,\,\,\Sigma^{i}=\left(
\begin{array}{cc}
\sigma^{i} & 0 \\
0 & \sigma^{i} \\	
\end{array}\right),\nonumber
\end{eqnarray}
where $I$ is the $2\times2$ identity matrix and $\vec{\Sigma}$, the spin vector. The matrices $\sigma^{i}$ $\left(\tau^{i}\right)$ are the standard Pauli matrices that satisfy the relation $\left(\sigma^{i}\,\sigma^{j}+\sigma^{j}\,\sigma^{i}\right)=2\,\eta^{ij}$ $(i,j,k=1,2,3)$. Furthermore, the $\gamma^{\mu}$ matrices are related to the $\gamma^{a}$ matrices as $\gamma^{\mu}=e^{\mu}_{\,\,\,a}\left(x\right)\gamma^{a}$ \cite{bd}. The representation of the Dirac matrices (\ref{8a}) is called the Weyl or chiral representation \cite{graf,Ja}.

Returning to Eq. (\ref{8}), we have that the term $\omega_{\mu ab}\left(x\right)$ obeys the Cartan structure equation \cite{md,ic,naka}
\begin{eqnarray}
d\hat{\theta}^{a}+\omega^{a}_{\,\,\,b}\wedge\hat{\theta}^{b}=0.
\label{9}
\end{eqnarray}
From Eq. (\ref{1.7}), we obtain the fllowing non-zero components of the connection 1-form $\omega^{a}_{\,\,\,b}$:
\begin{eqnarray}
\omega^{2}_{\,\,\,1}&=&-\omega^{1}_{\,\,\,2}=-\omega^{\,\,\,1}_{\varphi\,\,\,\,2}\left(x\right)\,d\varphi=\eta\,d\varphi,\nonumber\\
[-2mm]\label{10}\\[-2mm]
\omega^{3}_{\,\,\,1}&=&-\omega^{1}_{\,\,\,3}=-\omega^{\,\,\,1}_{\varphi\,\,\,\,3}\left(x\right)\,d\varphi=-\frac{n_{\Omega}}{2}\,d\varphi.\nonumber
\end{eqnarray}

In this way, we can write the Dirac equation for a massless fermion in this background as
\begin{eqnarray}
i\gamma^{0}\frac{\partial\Psi}{\partial t}+i\gamma^{1}\left(\frac{\partial}{\partial\rho}+\frac{1}{2\rho}\right)\Psi+i\frac{\gamma^{2}}{\eta\rho}\,\frac{\partial\Psi}{\partial\varphi}-i\gamma^{2}\frac{n_{\Omega}}{2\eta}\,\frac{\partial\Psi}{\partial y}+i\gamma^{3}\,\frac{\partial\Psi}{\partial y}-\frac{n_{\Omega}}{4\eta\rho}\,\gamma^{0}\gamma^{5}\,\Psi=0.
\label{11}
\end{eqnarray}

Since the extra coordinate is periodic, we can expand the Dirac spinor in the Fourier modes with respect to the extra dimension, whose scale is denoted by $L$:
\begin{eqnarray}
\Psi\left(t,\rho,\varphi,y\right)=\sum_{l=-\infty}^{\infty}\,e^{2\pi il\frac{y}{L}}\,\Psi_{l}\left(t,\rho,\varphi\right).
\label{12}
\end{eqnarray}  
Afterwards, the Dirac equation (\ref{11}) takes the form:
\begin{eqnarray}
i\gamma^{0}\frac{\partial\Psi_{l}}{\partial t}+i\gamma^{1}\left(\frac{\partial}{\partial\rho}+\frac{1}{2\rho}\right)\Psi_{l}+i\frac{\gamma^{2}}{\eta\rho}\,\frac{\partial\Psi_{l}}{\partial\varphi}+\frac{2\pi l}{L}\,\frac{n_{\Omega}}{2\eta}\,\gamma^{2}\,\Psi_{l}-\frac{2\pi l}{L}\,\gamma^{3}\,\Psi_{l}-\frac{n_{\Omega}}{4\eta\rho}\,\gamma^{0}\gamma^{5}\,\Psi_{l}=0.
\label{13}
\end{eqnarray}

We are interested in obtaing the geometric phase, therefore, we use the Dirac phase factor method \cite{dirac1,berry1}. Let us suppose that the Dirac spinor is written in the following form \cite{dirac1,berry1}
\begin{eqnarray}
\Psi_{l}\left(t,\rho,\varphi\right)=e^{i\phi}\,\Psi_{l}^{0}\left(t,\rho,\varphi\right),
\label{14}
\end{eqnarray}
where $\phi$ is the quantum phase which the wave function acquires in this quantum dynamics, and $\Psi_{l}^{0}$ is the solution of the Dirac equation
\begin{eqnarray}
i\gamma^{0}\frac{\partial\Psi_{l}^{0}}{\partial t}+i\gamma^{1}\,\frac{\partial\Psi_{l}^{0}}{\partial\rho}+i\frac{\gamma^{2}}{\eta\rho}\,\frac{\partial\Psi_{l}^{0}}{\partial\varphi}+\frac{2\pi l}{L}\,\frac{n_{\Omega}}{2\eta}\,\gamma^{2}\,\Psi_{l}^{0}-\frac{2\pi l}{L}\,\gamma^{3}\,\Psi_{l}^{0}=0.
\label{14a}
\end{eqnarray}
The last two terms in (\ref{14a}) correspond to the massive modes created by the extra-dimension of the Kaluza-Klein theory. Of course, by taking the mode $l=0$, we recover the description of massless fermions in graphene sheets in $\left(2+1\right)$ dimensions. Let us first discuss a general case where the fermions are massive. For massive fermions, we have that the spinor $\Psi_{l}$ acquires a geometric phase $\phi$ given by
\begin{eqnarray}
\phi&=&\oint\left(\frac{1}{2}\eta\,\Sigma^{3}-\frac{n_{\Omega}}{4}\,\Sigma^{2}\right)d\varphi\nonumber\\ 
[-2mm]\label{16.1}\\[-2mm]
&=&\pi\left(1-\frac{n_{\Omega}}{6}\right)\,\Sigma^{3}-\frac{n_{\Omega}}{2}\,\pi\,\Sigma^{2},\nonumber
\end{eqnarray}
where we have used the definition given in (\ref{eta}). The first contribution to the geometric phase (\ref{16.1}) comes from the polar coordinates, while the second and third contributions are due to the presence of the deficit angle of the topological defect.  We can see that the phase shift (\ref{16.1}) does not depend on the velocity of the quantum particle, then, the phase shift (\ref{16.1}) is a nondispersive phase \cite{disp,disp2,disp3}. Furthermore, we can see that we have obtained the phase shift (\ref{16.1}) without making the adiabatic approximation in a closed path, thus, the phase shift (\ref{16.1}) corresponds to a Aharonov-Anandan quantum phase \cite{ab1} for a massive fermion. 

From now on, let us concentrate on the zero mode ($l=0$) of the Kaluza-Klein expansion (\ref{14}), \textit{i.e.}, $\Psi_0\left(t,\rho,\varphi\right)\equiv\psi\left(t,\rho,\varphi\right)$, which describes the massless fermions in graphene in $\left(2+1\right)$ dimensions. In this case, the Dirac equation (\ref{11}) is reduced to
\begin{eqnarray}
i\,\gamma^{0}\,\frac{\partial\psi}{\partial t}+i\,\gamma^{1}\left(\partial_{\rho}+\frac{1}{2\rho}\right)\psi+i\,\frac{\gamma^{2}}{\eta\rho}\,\frac{\partial\psi}{\partial\varphi}-\frac{n_{\Omega}}{2\eta\rho}\,\gamma^{0}\gamma^{5}\,\psi=0.
\label{13}
\end{eqnarray}

By applying the Dirac phase factor method \cite{dirac1,berry1} given in (\ref{14}) for the zero mode ($l=0$), we have that the wave function of the massless fermion also acquires the same phase shift given in (\ref{16.1}). Therefore, the phase shift (\ref{16.1}) is also a nondispersive phase \cite{disp,disp2,disp3} and the Aharonov-Anandan quantum phase \cite{ab1} for a {\it massless fermion in graphene}. Then, the holonomy matrix (\ref{holonomy}) for a massless fermion in $\left(2+1\right)$ dimensions is given by
\begin{eqnarray}
\Psi'=U(C)\,\psi\left(t,\rho,\varphi\right)=\exp\left(i\left[\pi\left(1-\frac{n_{\Omega}}{6}\right)\,\Sigma^{3}-\frac{n_{\Omega}}{2}\,\pi\,\Sigma^{2}\right]\right)\,\psi\left(t,\rho,\varphi\right).
\label{holonomy2}
\end{eqnarray}

Note that the Dirac spinor $\psi$ in (\ref{13}) and (\ref{holonomy2}) is a $4$-component spinor even if the dimension is reduced to $\left(2+1\right)$ dimensions. By starting from the geometric description of the Fermi-points degree of freedom introduced by the Kaluza-Klein theory, we have that the $4\times4$ matrices $\Sigma^{i}$ that appears in (\ref{16.1}) and (\ref{holonomy2}) do not commute. The matrix $\Sigma^{3}$ corresponds to the generator of the rotation of the Dirac spinor on the $\rho\varphi$-plane, around the disclination cone. The contribution given by $\Sigma^{3}$ to the holonomy (\ref{holonomy2}) is similar to the holonomy (\ref{2.2b}) obtained via the parallel transport of a spinor around the apex of the cone. On the other hand, the matrix $\Sigma^{2}$ corresponds to the generator of rotations on the plane between the extra dimension and the radial direction, and its contribution to the holonomy (\ref{holonomy2}) is similar to the holonomy (\ref{2.2c}). This occurs due to the Kaluza-Klein description for graphene to be in $\left(3+1\right)$ dimensions. Therefore, The quantum phase acquired by the wave function of the massless fermion through the Kaluza-Klein theory is similar to the quantum phases obtained in Refs. \cite{lc1,lc2}.

\section{Conclusions}

We have shown a geometrical approach to describe the Fermi-point degrees of freedom and the phase shift acquired by the wave function of a massless fermion in graphene that arises from the presence of a disclination via the Kaluza-Klein theory. To develop this method, we have extended the two-dimensional metric with a topological defect used earlier (see Ref. \cite{fur}) by an extra compact dimension which describes the Fermi-Point degrees of freedom or the $K$-spin. We have shown that an Abelian gauge field can be introduced into the line element to describe the appearance of the quantum flux of the $K$-spin due to the influence of the topological defect on the $K$-spin part of graphene. Taking the zero mode of the Kaluza-Klein expansion of the wave function, we have found that the wave function of a massless fermion of the graphene acquires three independent contributions to the topological quantum phase, where one of them describes the phase shift which arises from the topology of the defect, and other describes the influence of the defect on the $K$-spin space. Moreover, by applying the Kaluza-Klein approach to graphene, we have seen that both topological quantum phases for massive modes and the massless mode are nondispersive, and can be obtained without making the adiabatic approximation in a closed path, corresponding to Aharonov-Anandan quantum phases.

We would like to thank the Brazilian agencies CNPq and CAPES for financial support. A. Yu. P. has been supported by the CNPq project No. 303461-2009/8.

\end{document}